\documentclass[prb,amssymb, amsmath,twocolumn,print]{revtex4}

\usepackage{pstricks}
\usepackage{amsmath}
\usepackage{graphicx,float}
\usepackage{graphics}
\usepackage{bm}

\newcommand{\ra}{\textsl{r}{^{*}}}

\newcommand{\rr}{\textbf{\textsl{r}}}
\newcommand{\ri}{\textsl{r}}

\begin{document}
\title [\textsf{Higher-order corrections to the relativistic perihelion advance\\ and the mass of binary pulsars.}]
{\textsf{Higher-order corrections to the relativistic perihelion advance\\ and the mass of binary pulsars.}}%
\author{\textsf{Maurizio M.} \surname{\textsf{D'Eliseo}}}%
\email{s.elmo@mail.com}
 \affiliation{Osservatorio S.Elmo \\ Via A.Caccavello 22,  80129 Napoli Italy}%

\begin{abstract}
We study the general relativistic orbital equation and using a
straightforward perturbation method and a mathematical device first
introduced by d'Alembert, we work out approximate expressions of a
bound planetary orbit in the form of trigonometrical polynomials and
the first three terms of the power series development of the
perihelion advance. The results are applied to a more precise
determination of the total mass of the double pulsar J0737-3039.
\end{abstract}
 \maketitle

\section{Introduction}

The general relativistic orbital equation for a planet revolving
around a star is deduced from the Schwarzschild line element
\begin{align}\nonumber
    ds^2&=c^2\gamma
    dt^2-\gamma^{-1}d\ri^2-\ri^2d\Omega^2,\\\label{Sc}
\gamma&=1-2\ra/\ri,\;\;\;d\Omega^2=d\theta^2-\sin^2\!\theta
\,d\varphi^2,
\end{align}
where $\ra=\mu/c^2$ and $\mu=GM$ are the gravitational radius and
the standard gravitational parameter of the star, respectively. All
remaining symbols have their usual meaning for this type of problem.
According to the geodesic hypothesis, the path followed by the
planet, considered as a test-body not to disturb the metric, can be
determined using the time-like Lagrangian
\begin{align}\label{lag}
2L&=\!\left(\frac{ds}{d\tau}\right)^{\!2}\!=c^2\gamma
\dot{t}^{\,2}\!-\gamma^{-1}\dot{\ri}^2\!
-\ri^2(\dot{\theta}^2\!+\sin^2\!\theta\,\dot{\varphi}^2),
\end{align}
and a variational principle that uses the functional
\begin{align}\label{sx}
S[q]=\int_{{\tau}_{1}}^{{\tau}_{2}}\!\!L(q,\dot{q})d\tau,\;\;\;
\dot{q}=dq/d\tau,
\end{align}
where $\tau$ is the planet's proper time and the function
$q=q(\tau)$ collectively denotes the degrees of freedom of a
possible generic planetary motion. The path actually followed is
what makes $S[q(\tau)]$ stationary, i.e. the functional derivative
\begin{align}\label{fd}
    \frac{\delta S[q(\tau')]}{\delta q(\tau)}=
    \int_{{\tau}_{1}}^{{\tau}_{2}}\frac{\delta L[q(\tau'),\dot{q}(\tau')]}{\delta
    q(\tau)}\,d\tau=0,
\end{align}
is zero when computed on the effective motion. From the working
point of view it is well known that a cancelation, in the integrand,
of the functional derivative is equivalent to that of the
Euler-Lagrange derivative
\begin{align}\label{fd1}
\frac{\delta }{\delta q}=\frac{\partial }{\partial
q}-\frac{d}{d\tau}\frac{\partial }{\partial\dot{q}},
\end{align}
and so we obtain four second-order differential equations (the
Euler-Lagrange equations) determining the sought time-like geodesic
\begin{align}\label{fd2}
\frac{\partial L}{\partial q}-\frac{d}{d\tau}\frac{\partial
L}{\partial\dot{q}}=0,\;\;\;
q=\ri,\theta,\varphi,t,\;\;\;\;\dot{q}=dq/d\tau.
\end{align}
It is worth noting that from the defining equation of the proper
time, $ds/d\tau=c$, we have $2L=c^2$, namely the Lagrangian is a
constant of the motion.

We observe that the Schwarzschild metric, and the Lagrangian, are
invariant under the reflection
\begin{align}\label{sym}
(t,\ri,\theta,\varphi)\mapsto(t,\ri,\pi-\theta,\varphi),
\end{align}
at the hyperplane $\theta=\pi/2$, so the mirror image of a geodesic
curve clearly has the same property. In particular, if we consider a
geodesic which at $\tau=0$ starts within the symmetry hyperplane and
is tangent to it, it must coincide with the transformed geodesic,
since the initial values of position and velocity determine a
geodesic uniquely. These considerations are confirmed by the
analysis of the Euler-Lagrange equation for $\theta$
\begin{align}\label{the}
\ddot{\theta}+\frac{2\dot{\ri}}{\ri}\dot{\theta}-\frac{\dot{\varphi}^2}{2}\sin2\theta=0,
\end{align}
which admits the solution $\theta=\pi/2$ satisfying the initial
conditions  $\theta_{0}=\pi/2$, $\dot{\theta}_{0}=0$. If we reorient
the coordinate system so that these conditions are met, the motion
takes place in the equatorial plane, and we can simplify the
Lagrangian~\eqref{lag} assuming $\sin\theta=1$, $\dot{\theta}=0$
\begin{align}\label{la}
2L=c^2\gamma
\dot{t}^{\,2}-\gamma^{-1}\dot{\ri}^2-\ri^2\dot{\varphi}^2.
\end{align}
 The coordinates $\varphi$ and $t$ are both cyclical: they appear in
the Lagrangian only in dotted form, and this means two conservation
laws. For the azimuth $\varphi$ we have
\begin{align}\label{el3}
-\frac{d}{d\tau}\frac{\partial
L}{\partial\dot{\varphi}}=0\Rightarrow-\frac{\partial
L}{\partial\dot{\varphi}}=\ri^2\dot{\varphi}=h=\text{const}.,
\end{align}
while for $t$ we have
\begin{align}\label{el4}
\frac{d}{d\tau}\frac{\partial
L}{\partial\dot{t}}=0\Rightarrow\frac{\partial
L}{\partial\dot{t}}=c^2\gamma \dot{t}=\kappa=\text{const}.\,.
\end{align}
The two constants are related to the angular momentum and to the
energy, respectively. Insertion of the two integrals into
Eq.~\eqref{la} leads to
\begin{align}\label{la1}
2L=\gamma^{-1}\frac{\kappa^2}{c^2}-\gamma^{-1}\dot{\ri}^2-\frac{h^2}{\ri^2}.
\end{align}
In place of the Euler-Lagrange equation for $\ri$, it is easier to
derive the radial equation from the integral $2L=c^2$, obtaining so
after multiplication by $\gamma$
\begin{align}\label{ri1}
\dot{\ri}^2+\frac{h^2}{\ri^2}-\frac{2\mu}{\ri}
-\frac{2h^2\ra}{\ri^3}+\left(c^2-\frac{\kappa^2}{c^2}\right)=0.
\end{align}
We need to cast the path equation into a form which clearly displays
the fact that we are dealing with a Keplerian orbit subjected to
small relativistic corrections, so we find it convenient to
eliminate the variable $\tau$ and introduce the angle $\varphi$
instead. This is possible since $\tau$ does not enter the equation
directly, but only via its differential $d\tau$, so we transform the
$\tau$-derivative into a $\varphi$-derivative, according to the
identity $d/d\tau=(h/\ri^2)d/d\varphi$. Moreover, the formulas
simplify considerably if we replace $\ri$ by its reciprocal
$u=1/\ri$. Denoting by a prime differentiation with respect to
$\varphi$, we have $\dot{\ri}^2=h^2u'^2$ and Eq.~\eqref{ri1} becomes
\begin{align}\label{ri2}
u'^2+u^2-\frac{2\mu}{h^2}u-2\ra
u^3+\frac{1}{h^2}\left(c^2-\frac{\kappa^2}{c^2}\right)=0,
\end{align}
while with a differentiation with respect to $\varphi$ we find
\begin{align}\label{ri3}
u'\left(u''+u-\frac{\mu}{h^2}-3\ra u^2\right)=0.
\end{align}
A solution of this equation is $u'=0$, or $u=\text{const.}$, and
this means a circular orbit. Ruling out this possibility,
Eq.~\eqref{ri3} requires that
\begin{align}\label{b}
u''+u=\frac{\mu}{h^2}+3\ra u^2.
\end{align}
Comparing this equation with the classical Binet's formula for a
particle subjected to a central force of magnitude $f(u)$ in polar
coordinates
\begin{align}\label{b1}
u''+u=-\frac{f(u)}{h^2u^2},
\end{align}
it appears that a particle moving in the Schwarzschild field will
behave as though it were under the influence of an effective
Newtonian inverse-square force plus an additional fourth-power
inverse force
\begin{align}\label{fo}
-\left(\frac{\mu}{\ri^2}+\frac{3\ra
h^2}{\ri^4}\right)\!\frac{\rr}{\ri},
\end{align}
in a framework in which proper time is used as independent variable.
We remark that for planetary trajectories, in Eq.~\eqref{b} the
terms $u$ and $\mu/h^2$ are comparable, while $u$ and $3\ra u^2$
differ by a factor $3\ra u$. The maximum value of this quantity
corresponds to the planet which is nearest to the star, so, for
Mercury, with
$\ra\approx1.476\cdot10^{5}\text{cm},\;\ri\approx5.5\cdot10^{12}$cm,
it is $3\ra/\ri\approx10^{-7}$. This implies that Eq.~\eqref{b}
represents in general an oscillator with a constant forcing term and
a weak quadratic nonlinearity which affects its
frequency\citep{nay73}.

The orbital equation~\eqref{b} is not the complete solution to
motion problem, which would require the knowledge of the function
$\varphi(t)$, being $t$ the coordinate time, which is the time
measured by an observer at rest at great distance from the origin,
and therefore the quadrature of the equations
\begin{align}
\frac{d\tau}{d\varphi}=\frac{1}{hu^2},\;\;\;\;
\frac{dt}{d\varphi}=\frac{\kappa}{\gamma hu^2},
\end{align}
but this does not concern us here.

The importance of Eq.~\eqref{b} is due to a variety of reasons. It
originates, without any approximation, from an exact solution of
Einstein's equation. Because of the lacking of a time variable, it
is entirely geometric, so it can be employed to deduce the precise
shape of the orbit and one of the most important post-Newtonian
predictions of the theory: the advance of the perihelion of an
elliptical orbit \citep{din92}. The equation, then, is naturally
linked to Newtonian dynamics, that in the description of a planetary
motion provides already an excellent approximation as well as a
clear connection with observation. Last, it is simple when compared
with other post-Newtonian equations of planetary dynamics, and
therefore its mathematical and theoretical limits can be clearly
defined. We will use it to calculate the orbit of a test particle in
the gravitational field external to a non-spinning spherical mass
eventually to the order $1/c^n$ for any arbitrary positive integer
$n$, and the corresponding formulas of the periastron advance to the
same order. If the effects predicted fall in the range of the
observability, measurability or indirect determination for those
physical systems where the equation is applicable, then no doubt
they should necessarily be taken into account in all specific cases.

A determination of higher-order terms of the periastron advance of a
binary pulsar by using the second post-Newtonian (2PN) method  have
been effected by some auctors \citep{dam88,sch93}. Their results
based on the 2PN theory can be applied to the problem of a test
particle under a strong gravitational field by letting the mass of
one pulsar theoretically approach zero. Since one can argue about
the rigor of this and other methods used to handle the 2PN problem
of motion, we think that a proper analysis of the correctness and of
the limits of accuracy of these approaches should be based on the
agreement under some convenient limit with the results exposed in
the present paper. When considering relativistic effects, other
post-Keplerian phenomena come into play in the motion of bodies. For
example, one could  insert the relativistic force~\eqref{fo} as a
perturbing radial acceleration in Gauss equations for the variations
of the Keplerian orbital elements  \citep{ber03}. In particular, in
the expression of the time derivative of the mean anomaly, in
addition to the radial force it appears explicitly the motion of the
perihelion. We could cite at this regard two recent works
\citep{ior07,lin10}on the secular advance of the mean anomaly in
binary systems, which could be easily extended to include the
higher-order effects calculated in this paper. It would also be
interesting to compare the results so obtained with those computed
by means of the post-Newtonian Lagrangian planetary
equations\citep{cal97}, and this could be the subject of future
work.

\section{The iteration method}

In the following we will express $h$ in terms of elliptic elements
of Newtonian approximation, so $h=\sqrt{\mu p}$, where $p=a(1-e^2)$
is the semi-latus rectum, $a$ is the semi-major axis and $0\leq e<1$
is the eccentricity\citep{del07}. Although we have excluded before
the circular orbit, the orbital equation encompasses also this
possibility. To make its structure more apparent and to facilitate
the calculations, we cast Eq.~\eqref{b} in dimensionless form, which
represents a well-known problem in mathematical physics
\begin{align}\label{d1}
u''+u=1+\epsilon u^2,
\end{align}
where this time $u$ means $p/\ri$ and $\epsilon$ is the pure number
$3\ra/p$. Thus, when $\ri=a(1\pm e)$, $u=1\mp e$. To find an
approximate solution to this equation, one could use regular
perturbation theory, writing $u$ in the form of a perturbative
expansion in powers of $\epsilon$, but it runs in trouble here,
since we wish an expansion that converges for all values of the
independent variable $\varphi$. To do this in the most convenient
way, we perform a shift in the origin $w=u-1$, in order to write the
equation as that of the perturbed harmonic oscillator
\begin{align}\label{d0}
w''+w=\epsilon(1+w)^2.
\end{align}
When $\epsilon=0$ the equation becomes
\begin{align}\label{d2}
w''_{0}+w_{0}=0,
\end{align}
whose general solution is
\begin{align}\label{di3}
w_{0}(\varphi)=w_{0}(0)\cos\varphi+w'_{0}(0)\sin\varphi.
\end{align}
We shall adopt the initial conditions $w_{0}(0)=e,\,w'_{0}(0)=0$,
which hereafter we shall denote standard, and so
\begin{align}\label{d3}
w_{0}(\varphi)=e\cos\varphi.
\end{align}
This choice of the initial conditions leads to a simplification of
the algebra. Solution~\eqref{d3} has frequency one and period
$2\pi$. The equation of the orbit is then
\begin{align}\label{ell}
\frac{w_{0}+1}{p}=\frac{1}{\ri}=\frac{1+e\cos\varphi}{p},
\end{align}
that is an ellipse, where the angle $\varphi=0$ locates the position
of the perihelion (because there the function $w_{0}$ is maximal),
and identifies the direction of the apse line, the greatest symmetry
axis of the orbit in the plane.

If we try a straightforward iterative perturbation scheme to solve
Eq.~\eqref{d0} attaching the appropriate subscripts (the iteration
numbers) to its sides, for the first-order solution $w_{1}$ we
obtain the equation
\begin{align}\label{}
w''_{1}+w_{1}&=\epsilon(1+w_{0})^2,
\end{align}
of a forced pendulum where there is a resonant $\cos\varphi$ term.
If we do not want that the pendulum oscillates with an ever
increasing period ($w_{1}$ must stay small for all values of
$\varphi$), then the external force is not allowed to have a Fourier
component with the same periodicity as the pendulum itself. Note
here and in the following that, according to the method of
undetermined coefficients, the particular solution of an equation of
the form
\begin{align}\label{gs}
w''+k^2w=\sum_{n}X_{n}\cos nk\varphi,
\end{align}
is formally expressed by
\begin{align}\label{gs1}
w=\frac{X_{0}}{k^2}+\left[\frac{X_{n}\cos
nk\varphi}{k^2(1-n^2)}\right]_{n=1}+\sum_{n>1}\frac{X_{n}\cos
nk\varphi}{k^2(1-n^2)},
\end{align}
whose second term is singular, but can be regularized since,
applying the L'Hospital's rule, we have
\begin{align}\label{hr}
\lim_{n\rightarrow1}\frac{X_{n}\cos
nk\varphi}{k^2(1-n^2)}=\frac{X_{1}\varphi\sin k\varphi}{2k},
\end{align}
with $\varphi$ explicitly present outside the argument of the
trigonometrical function, and therefore growing without limits with
time $t$, since $\varphi$ is a monotonic function of $t$. Obviously
this would destroy the stability of the orbit. Therefore this simple
perturbation scheme does not work for Eq.~\eqref{d0}. Things go
differently if we rearrange Eq.~\eqref{d0} in the form
\begin{align}\label{d00}
&w''+k_{\epsilon}^2w=\epsilon(1+w^2),&k_{\epsilon}^2\equiv
1-2\epsilon.
\end{align}
The equation is unchanged, but now we have chosen to look at the
isolated linear term on the right side as part of the unperturbed
equation, which is
\begin{align}\label{d001}
w''_{0}+k_{\epsilon}^2w_{0}=0.
\end{align}
Equation~\eqref{d001} now represents an oscillator whose natural
frequency is $k_{\epsilon}$, smaller than 1, and its solution
obeying to the standard initial conditions is
\begin{align}\label{hom}
w_{0}(\varphi)=e\cos k_{\epsilon}\varphi.
\end{align}
It follows that, with respect to the Newtonian
oscillator~\eqref{d2}, in the relativistic case, to the lowest
order, the values of $\ri$, which trace out an approximated ellipse,
do not begin to repeat until somewhat after the radius vector has
made a complete revolution. Hence the orbit may be regarded as being
an ellipse which is slowly rotating. In particular, the angular
advance for revolution of the apse line is given by
\begin{align}\label{dt}
\Delta\omega=2\pi\left(\frac{1}{k_{\epsilon}}-1\right)=2\pi\epsilon+O(\epsilon^2).
\end{align}
This is Einstein's perihelion formula, but it represents only the
first term of a series development in powers of $\epsilon$. Our aim
is to compute this series up to order $\epsilon^3$. Although the
first approximation gives a satisfactory degree of accuracy for
ordinary planetary problems, and the second one can cope with
particular astrophysical situations, we push a step further the
computation because with little extra work we shall exhaust all
conceivable theoretical needs before the orbital equation fails to
represent the relativistic motion of bodies in strong-field
situations.

Let us try again the perturbation scheme on Eq.~\eqref{d00}. We have
now
\begin{align}\label{d0p}
w''_{1}+k_{\epsilon}^2w_{1}&=\epsilon(1+w^2_{0}),\\
&=\epsilon\left(1+\frac{1}{2} e^2\right)+\epsilon\frac{1}{2} e^2\cos
2k_{\epsilon}\varphi,
\end{align}
that we solve with the standard initial conditions. It is to be
noticed that this time we have not any more the resonant term $\cos
k_{\epsilon}\varphi$. This equation is of the type
\begin{align}\label{ge}
w''_{1}+k_{\epsilon}^2w_{1}=A+B\cos 2k_{\epsilon}t.
\end{align}
The solution is the sum of the general solution of the homogeneous
part and of a particular integral of the complete equation. This
integral can be found using formula~\eqref{gs1}. Thus the solution
of Eq.~\eqref{ge} is
\begin{align}
w_{1}=E\cos
k_{\epsilon}\varphi+Ak_{\epsilon}^{-2}-\frac{Bk_{\epsilon}^{-2}}{3}\cos2k_{\epsilon}\varphi.
\end{align}
As $w_{1}$ is of order $\epsilon$ as $A,B$ are, we put
$k_{\epsilon}^{-2}=1$, while the constant $E$ is determined by the
standard initial conditions to be
\begin{equation}
E=e-\epsilon\left(1+\frac{e^2}{3}\right),
\end{equation}
and so we find
\begin{align}\nonumber
w_{1}=\epsilon\left(1+\frac{1}{2}e^2\right)&+\left[e-\epsilon\left(1+\frac{1}{3}e^2\right)\right]\!\cos
k_{\epsilon}\varphi\\\label{x1}&-\epsilon\frac{1}{6} e^2\cos
2k_{\epsilon}\varphi,
\end{align}
which we shall write in abridged form
\begin{align}\label{af}
w_{1}= A_{1}+(e+E_{1})\cos k_{\epsilon}\varphi+B_{1}\cos
2k_{\epsilon}\varphi,
\end{align}
where as a notational aid we agree, here and in the following, that
the subscript number $i$ attached to a capital letter representing a
coefficient wants to emphasize the presence of the factor
$\epsilon^i$.
 This solution is periodical and bounded for all values of $\varphi$, and
represents the general relativistic first-order deviation from the
classical elliptical orbit. In the next iteration, which should give
the solution correct to order $\epsilon^2$
\begin{align}\label{d0p1}
w''_{2}+k_{\epsilon}^2w_{2}&=\epsilon(1+w^2_{1}),
\end{align}
the right-hand side will present again a resonant term, since to
this order we get
\begin{align}\label{abch}
&w''_{2}+k_{\epsilon}^2w_{2}=A+H_{2}\cos k_{\epsilon}\varphi+B\cos2k_{\epsilon}\varphi+C\cos3k_{\epsilon}\varphi,\\
\label{abc}&\text{where}\;\;\; H_{2}\equiv\epsilon e(2A_{1}+
B_{1})=\epsilon^2\frac{e(12+5e^2)}{6},
\end{align}
with $A,B,C$ to be written later. To get rid of the resonant term
$H_{2}\cos k_{\epsilon}\varphi$ on the right side of
Eq.~\eqref{abch} we will use a device which stems naturally from the
logical path we followed in writing Eq.~\eqref{d00} and that was
introduced for the first time by d'Alembert to control the plague of
the secular terms present in the integration of the equation of the
lunar motion, which is of the same type of that we are considering
here \citep{dal54}. In order to suppress the unwanted cosine term,
we add a counter term and write Eq.~\eqref{d00} in the form
\begin{align}\label{d002}
w''+\left(k_{\epsilon}^2-\frac{H_{2}}{e}\right)w=\epsilon(1+w^2)-\frac{H_{2}}{e}w,
\end{align}
and consequently we consider the approximate equation
\begin{align}\label{d0p1a}
w''_{2}+k_{\epsilon^{\!2}}^2
w_{2}&=\epsilon(1+w^2_{1})-\frac{H_{2}}{e}w_{1},\\\label{d0p1b}
k_{\epsilon^{\!2}}^2&\equiv k^2_{\epsilon}-\frac{H_{2}}{e}.
\end{align}
In the right side, the added term suppresses the resonant term,
since replacing $w_{1}$ with $e\cos k_{\epsilon^{\!2}}\varphi$ (the
other terms of $w_{1}$ would give terms of order greater than
$\epsilon^2$), we obtain just $-H_{2} \cos
k_{\epsilon^{\!2}}\varphi$. In the left side the coefficient
$k_{\epsilon}^2$ will be diminished by the amount $H_{2}/e$, so that
we get
\begin{align}\label{ke}
k_{\epsilon^{\!2}}^2=1-2\epsilon-\frac{12+5e^2}{6}\epsilon^2.
\end{align}
This way we shall obtain an acceptable solution $w_{2}$ of
Eq.~\eqref{d0p1} to order $\epsilon^2$ and, at the same time, the
correction to the frequency to the same order. The procedure can
obviously be repeated until we have reached the required
approximation degree for both solution and frequency. The essence of
this method was rediscovered by Lindstedt \citep{lin83} and its
practical application was further elaborated by Poincar\'{e}
\citep{nay73}. Equation~\eqref{d0p1a} has the form
\begin{align}\label{abcd}
w''_{2}+k_{\epsilon^{\!2}}^2w_{2}&=A+B\cos2k_{\epsilon^{\!2}}\varphi+C\cos3k_{\epsilon^{\!2}}\varphi,\\
A&\equiv\frac{(2+e^2)}{2}\epsilon-\frac{(3e+e^3)}{3}\epsilon^2,\\
B&\equiv\frac{e^2}{2}\epsilon-\frac{3e+e^3}{3}\epsilon^2,\\
C&\equiv-\frac{e^3}{6}\epsilon^2,
\end{align}
and its general solution, by Eq.~\eqref{gs1}, is
\begin{align}\nonumber
w_{2}=E\cos
k_{\epsilon^{\!2}}\varphi&+Ak_{\epsilon^{\!2}}^{-2}-\frac{Bk_{\epsilon^{\!2}}^{-2}}{3}\cos
2k_{\epsilon^{\!2}}\varphi\\\label{solu}&-\frac{Ck_{\epsilon^{\!2}}^{-2}}{8}\cos
3k_{\epsilon^{\!2}}\varphi,
\end{align}
where now it suffices to put $k_{\epsilon^{\!2}}^{-2}=1+2\epsilon$.
We thus find, by determining $E$ by means of the standard initial
conditions, the second-order approximation to the orbit
\begin{align}\nonumber
w_{2}=A_{1}&+A_{2}+(e+E_{1}+E_{2})\cos
k_{\epsilon^{\!2}}\varphi\\\label{w2} &+(B_{1}+B_{2})\cos
2k_{\epsilon^{\!2}}\varphi+C_{2}\cos3k_{\epsilon^{\!2}}\varphi,
\end{align}
where
\begin{align*}
A_{2}&=\epsilon^2\frac{6-3e+3e^2-e^3}{3},\\
E_{2}&=\epsilon^2\frac{29e^3-96e^2+96e-288}{144},\\
B_{2}&=\epsilon^2\frac{e^3-3e^2+3e}{9},\\
C_{2}&=\epsilon^2\frac{e^3}{48}.
\end{align*}
Let us consider now the next equation
\begin{align}\label{d0d}
w_{3}''+k_{\epsilon^{\!2}}^2w_{3}=\epsilon(1+w_{2}^2)-\frac{H_{2}}{e}w_{2}.
\end{align}
Since we are interested only in the resonant term of $O(\epsilon^3)$
we do not solve this equation, but we can extract quickly from the
right side the secular generating term $H_{3}\cos
k_{\epsilon^{\!2}}\varphi$  (see Appendix), where
\begin{align}\label{ee3}
H_{3}=\epsilon e(2A_{2}+B_{2})
=\epsilon^3\frac{e(36-15e+15e^2-5e^3)}{9}.
\end{align}
Following d'Alembert's method, this term will be canceled writing
Eq.~\eqref{d0d} in the form
\begin{align}\label{d0d1}
w_{3}''+k_{\epsilon^{\!3}}^2w_{3}&=\epsilon(1+w_{2}^2)
\!-\!\frac{H_{2}+H_{3}}{e}w_{2},\;\;k_{\epsilon^{\!3}}^2\!=k_{\epsilon^{\!2}}^2-\frac{H_{3}}{e},
\end{align}
and we have
\begin{align}\label{k3}
k_{\epsilon^{\!3}}^2=1-2\epsilon-\frac{12+5e^2}{6}\epsilon^2-\frac{36-15e+15e^2-5e^3}{9}\epsilon^3.
\end{align}
Further, we find
\begin{align}
k_{\epsilon^{\!3}}&=1-\epsilon-\frac{18+5e^2}{12}\epsilon^2-\frac{126-30e+45e^2\!-10e^3}{36}\epsilon^3,\\
k_{\epsilon^{\!3}}^{-1}&=1+\epsilon+\frac{30+5e^2}{12}\epsilon^2+\frac{270-30e+75e^2\!-10e^3}{36}\epsilon^3.
\end{align}
The rotation for revolution of the periapse, by denoting with
$\omega$ its angular measure, is given, to order $\epsilon^3$, by
\begin{align}\nonumber
\Delta\omega&=2\pi\left(k_{\epsilon^{\!3}}^{-1}-1\right)
=2\pi\epsilon\!+5\pi\!\left(\!1\!+\frac{1}{6}e^2\!\right)\!\epsilon^2
\!\\\label{dom}&\phantom{=}\qquad\;\;\;\;\;+5\pi\!\left(3-\frac{1}{3}e+\frac{5}{6}e^2\!-\frac{1}{9}e^3\!\right)\!\epsilon^3.
\end{align}
This formula agrees with the results obtained with the
Poincar\'{e}-Lindstedt method and some of its equivalent
modifications \citep{mas84,don98}. By denoting with $P$ the
anomalistic period, that is the time that elapses between two
passages of the object at its perihelion expressed in days, the
average advance rate is
\begin{align}\nonumber
\dot{\omega}(\text{rad/d})&=+\frac{\Delta\omega}{P}\approx
\dot{\omega}_{1}+\dot{\omega}_{2}+\dot{\omega}_{3}\\\nonumber
 &=+\frac{6\pi\ra}{a(1-e^2)P}
+\frac{15\pi\ra^2(6+e^2)}{2a^2(1-e^2)^2P}\\\label{ei}
&\phantom{=}\;+\frac{15\pi{\ra}^3(54-6e+15e^2-2e^3)}{2a^3(1-e^2)^3P}.
\end{align}
What is the meaning of $e$ in the solution $w_{n}$ of order
$\epsilon^n$? In the zero-order solution $w_{0}$, $e$ is the
eccentricity, but in the successive approximations it loses this
characterization: the symbol $e$ is simply a constant in the open
interval (0,1) that one introduces in the initial conditions to
express the shape of the orbit, and that coincides with the
eccentricity of the osculating Kepler ellipse to the path followed
by the planet when $\varphi=0$.

\section{Applications}

 In astrophysical
applications Eq.~\eqref{ei}, written in the form
\begin{align}\label{algi}
f(\ra, a, e,\Delta\omega)=0,
\end{align}
is an implicit relation between dynamical and orbital parameters
characterizing the system under consideration, and it can be used to
calculate any of them, once known the others. Thus, for example, in
the solar system $\ra, a, e$ are known, and we calculate
$\Delta\omega$, the perihelion shift of the planetary orbits. It is
also evident that a periastron advance is highly enhanced by small
$a$'s (whence short periods) and high orbital eccentricities. A word
of caution is needed here: we must consider the fact that a very
large eccentricity can also mean a very small periastron distance
$a(1-e)$, and so we must stop precisely at the orbit that just
grazes the surface of the star. The problem of detecting the motion
of periastron (or of apoastron) of some highly elliptic extrasolar
planets has been considered by some authors \citep{ior06,pal08}, and
of course now the question is to determine the magnitude of the
higher orders effects for some plausible orbital parameters. In the
first column of Table I for comparison purposes we have inserted the
data concerning the Sun-Mercury system, while the second and third
columns are referred to two hypothetical exoplanets of a solar-mass
star with great eccentricities and/or small radial distances.
\begin{table}
\caption{Mercury and two hypothetical exoplanets}
\begin{tabular}{|c|c|c|c|}
 \hline
   System$\;\rightarrow$        &Sun/Mercury&Star/Alpha&Star/Beta\\
 \hline
 \hline
$M_{\odot}$                           &1.00                     & 1.00                       &1.00                       \\
$\alpha\,(\text{cm}) $                &$1.475\cdot10^5 $        & $1.475\cdot10^5 $           &$1.475\cdot10^5 $                        \\
$a\,(\text{cm})$                      &$5.791\cdot10^{12}$      & $5.791\cdot10^{12}$         &$8.788\cdot10^{10}$                        \\
$e$                                   &0.2056                   & 0.95                         &0.20                                    \\
$\epsilon$                            &$1.038\cdot10^{-5}$      &$1.038\cdot10^{-5}$          &$1.038\cdot10^{-5}$\\
$P\,(\text{d})$                       &87.9                     & 87.9                       &0.164                                    \\
\hline
$\dot{\omega}_{1}\,(\text{rad/d})$    &$5.703  \cdot  10^{-9}$  & $5.602  \cdot  10^{-8}$     &$2.001  \cdot  10^{-4}$                  \\
$\dot{\omega}_{2}\,(\text{rad/d})$    &$1.097  \cdot  10^{-15}$ & $1.262  \cdot  10^{-13}$      &$2.652  \cdot  10^{-9}$                  \\
$\dot{\omega}_{3}\,(\text{rad/d})$    &$2.46  \cdot  10^{-22}$  & $2.873  \cdot  10^{-19}$     &$4.098  \cdot  10^{-14}$                 \\
\hline
$\Delta\omega_{1}\,(\text{arcsec/yr})$   &$0.429$                & $4.220$                   & $1.514  \cdot  10^{4}$                                   \\
$\Delta\omega_{2}\,(\text{arcsec/yr})$ &$2.26 \cdot  10^{-10}$   &$9.51 \cdot  10^{-6}$        & $1.998 \cdot  10^{-1}$\\
$\Delta\omega_{3}\,(\text{arcsec/yr})$ &$5.09 \cdot  10^{-17}$   &$2.16 \cdot  10^{-11}$      & $3.088  \cdot  10^{-6}$ \\
 \hline
\end{tabular}
\end{table}
While is doubtful, given the particularly high levels of
observational accuracy required, that it is actually possible to
find planets with orbital characteristics fitted for this purpose,
in meantime one can imagine a verification of the high-orders
perihelion formula achieved by means of a man-made solar probe in a
carefully planned celestial mechanics experiment. However, we
believe that the higher order terms in Eq.~\eqref{ei} must be taken
into account in all attempts to detect the Sun's Lense-Thirring
effect on the perihelia of the inner planets in order to separate
with improved accuracy these two important consequences of general
relativity.

 Some close binary systems are good
subjects for a verification and an application of the formulas
\citep{kra06}. Before applying Eqs.~\eqref{dom}, \eqref{ei} to a
concrete example, we recall that a generalization of Enstein's
perihelion formula says that in a system of two non negligible
spherical masses, the two-body problem, the gravitational parameter
$\mu$ means $GM=G(m_{1}+m_{2})$ \citep{lan66}. In this instance
$\ra$ could be loosely named the equivalent or the nominal
gravitational radius of the system. This result is only the
first-order approximation to a full relativistic treatment of the
center of mass of two comparable bodies, a hitherto unresolved
problem, but however we will fully exploit the effect of this
approximation. For these systems, the semimajor axis and the total
mass $M$ are not directly determinable, while eccentricities,
periods and periastron advance rate are, so we can use
Eq.~\eqref{algi} to find $\ra$, once we have replaced $a$ with the
period $P$ and the total mass $M$ (the degeneracy of the mass is
removed by measurement of another relativistic orbital
characteristic, the Einstein parameter) using Kepler's third law in
the form
\begin{align}\label{k31}
a=\left(\frac{P^2GM}{4\pi^2}\right)^{1/3}=\left(\frac{Pc}{2\pi}\right)^{2/3}\ra^{1/3}.
\end{align}
To be precise, this law is not exactly true in a regime of strong
fields and little distances, but it can be used harmlessly in a
perturbative calculation for the systems we are considering within
the same constraints we have set for the center of mass, that is a
first-order approximation. Incidentally, this will free our results
from other complex and still poor understood effects and, besides,
this is in line with the regime of validity of the equation of
motion which we have supposed. From Eqs.~\eqref{ei},~\eqref{k31} we
get
\begin{align}\nonumber
\dot{\omega}=&\left(\frac{2\pi}{P}\right)^{\!5/3}\left(\frac{\ra}{c}\right)^{2/3}f(e)+\left(\frac{2\pi}{P}\right)^{\!7/3}\left(\frac{\ra}{c}\right)^{4/3}g(e)\\
\label{mio}&\qquad+\left(\frac{2\pi}{P}\right)^{3}\left(\frac{\ra}{c}\right)^{2}h(e),
\end{align}
where
\begin{align*}
f(e)&=\frac{3}{(1-e^2)},\\
g(e)&=\frac{15(6+e^2)}{2(1-e^2)^2},\\
h(e)&=\frac{15(54-6e+15e^2-2e^3)}{4(1-e^2)^3}.
\end{align*}
To obtain directly the value of the mass $M$ in units of the solar
mass $M_{\odot}$, we can replace in Eq.~\eqref{mio} the ratio
$\ra/c$ with the product $T_{\odot}M$, where $T_{\odot}\equiv
GM_{\odot}/c^3$ is the mass of the Sun expressed in units of time.
In the current literature the expression of $\dot{\omega}$ lacks of
the last two term of Eq.~\eqref{mio}. \citep{kra06} Inserting the
values of $e,P,\dot{\omega}$ determined for a given binary system,
Eqs.~\eqref{mio},~\eqref{ei} can be numerically solved for $\ra$ and
$M$ respectively, and will give the value of the gravitational
radius of the system or the total mass. Once known $\ra$, we can
complete the calculation and use Eq.~\eqref{ei} to compute the value
of $a$. The value of $\dot{\omega}$ that one determines is due, in
general, to relativity plus extra classical terms, as  the
gravitational quadrupole moment induced by rotation and tidal
deformations.  But strongly self-gravitating objects as binary
pulsars have a mass pointlike behavior, and thus the motion of their
periastron must be entirely ascribed to relativity. Strictly
speaking, since this bodies are rapidly spinning, one should
consider also the Lense-Thirring effect or use the Kerr solution to
the Einstein equation, but this is an argument for a further study.

We apply the theory to the double pulsar J0737-3039. Its orbital
period is the smallest so far known for such an object, and it can
be determined with great precision along with $\dot{\omega}$. For a
such system cumulative effects add rapidly, and this allows a
meaningful application of the formulas, despite the rather small
eccentricity.
\begin{table}[h]
\caption{Pulsar J0737-3039}
\begin{tabular}{@{}|ll|}
  \hline
$e$ & 0.0877775                    \\
$P$ &$ 0.10225156248\text{(d)}\;\sim\;8834.534991$(s)  \\
$\dot{\omega}$  &$ 16.89947^{o}(\text{yr}^{-1})\;\sim\;9.346445651\cdot10^{-9}(\text{s}^{-1})$  \\
\hline \hline
$c$ & $2.99 792 458\cdot10^{10} $(cm/s)  \\
$T_{\odot}$ &$4.925490947\cdot10^{-6}$(s) \\
$yr$ &$ 3.15576 \cdot  10^{7}$(s)    \\
\hline
\end{tabular}
\end{table}
\begin{table}[h]
\caption{Pulsar J0737-3039 derived parameters}
\begin{tabular}{@{}|lll|}
  \hline
To order:&$\qquad\epsilon$&$\qquad\epsilon^3$\\
 \hline
$\ra\,(\text{cm}) $   & $3.82014\cdot10^5 $           &$3.8199525\cdot10^5 $        \\
$M(\text{M}_{\odot})$            & 2.587075                       &$2.586948$              \\
$a\,(\text{cm})$         & $8.788391\cdot10^{10}$         &$8.788680\cdot10^{10}$           \\
$\epsilon$               &$1.314166\cdot10^{-5}$      &$1.314057\cdot10^{-5}$\\
\hline \hline
$\dot{\omega}_{1}$&$16.89947^{o}\text{yr}^{-1}$ &$16.89891408^{o}    \text{yr}^{-1}$     \\
$\dot{\omega}_{2}$&- &$\;\, 0.00055589^{o}\text{yr}^{-1}$   \\
$\dot{\omega}_{3}$&- &$\;\, 0.00000002^{o}\text{yr}^{-1}$  \\
\hline
$\dot{\omega}$    &$16.89947^{o}\text{yr}^{-1}$ &$16.89946999^{o}    \text{yr}^{-1}$ \\
\hline
\end{tabular}
\end{table}
In Table II we have indicated the determined \citep{kra06} values of
$e,P,\dot{\omega}$ and, for the reader's convenience, some numerical
values used in the calculations, while in Table III are indicated
all parameters that can be derived through an application of our
formulas.
 In the first column, the computation are done to order
 $\epsilon$, assuming the validity of general relativity
thorough the use of Einstein's precession formula $\dot{\omega}_{1}$
of Eq.~\eqref{ei},  while in the second column are inserted the
values we have computed to order $\epsilon^3$, in particular how
much of $\dot{\omega}$ comes from $\dot{\omega}_{i},\;i=1,2,3$.
Since from the observational point of view the single higher-orders
contributions to the precession rate are inextricably combined, and
so hidden to a direct measurement, we can obtain an indirect
verification turning to the main consequence of this approximation:
a diminution by a small amount of the total mass of the system with
respect to the currently accepted value, with the consequent
redefinition of other parameters, in particular of $a$, and so we
can refine the model of the system with one that fits the found
variations.

Ever more accurate determinations of $\dot{\omega}$ might enable,
for this as for any other similar system, to test the usefulness of
the third-order approximation to the periastron secular motion
deduced from the equation of motion within the limits of
approximation that we have introduced. However, the results found
can be considered strictly true for the motion of a test body in the
gravitational field of a central body of mass equivalent to the
total mass of the binary system.

\section{Appendix}

By simple considerations of powers and arguments, and with the aid
of the subscript notation, we can quickly find the coefficient of
the secular term $H_{n}$ once known $H_{2},\dots,H_{n-1}$ and the
orbit to order $\epsilon^{n-1}$.  Here is a sketch of how we unearth
from the right side of Eq.~\eqref{d0d}
\begin{align}\label{did}
    \epsilon(1+w_{2}^2)-\frac{H_{2}}{e}w_{2}=\epsilon(1+w_{2}^2)-\epsilon(2A_{1}+B_{1})w_{2},
\end{align}
the resonant term $H_{3}\cos k_{\epsilon^{\!2}}\varphi$ of order
$\epsilon^3$. \\We consider first the multinomial expression
\begin{align}\nonumber
\epsilon w_{2}^2=\epsilon[A_{1}&+A_{2}+(e+E_{1}+E_{2})\cos
k_{\epsilon^{\!2}}\varphi\\\label{ep3} &+(B_{1}+B_{2})\cos
2k_{\epsilon^{\!2}}\varphi+C_{2}\cos3k_{\epsilon^{\!2}}\varphi]^2,
\end{align}
together with the following algebraic and trigonometrical identities
\begin{align}\nonumber
&\epsilon(a+b+c+\dots)^2=\epsilon a^2+\epsilon b^2+\epsilon c^2+\dots\\
\nonumber&\phantom{\epsilon(a+b+c+\dots)^2=}\;\dots+2\epsilon ab+2\epsilon ac+\dots\\
&\label{mu1}\phantom{\epsilon(a+b+c+\dots)^2=}\;\dots+2\epsilon bc+\dots,
\\
&\label{mu2}(\cos nk_{\epsilon^{\!2}}\varphi)^2=\frac{1}{2}\cos 2nk_{\epsilon^{\!2}}\varphi+\frac{1}{2},\\
&\label{mu3}a\cos(n+1)k_{\epsilon^{\!2}}\varphi\cdot b\cos
nk_{\epsilon^{\!2}}\varphi=\frac{ab}{2}\cos
k_{\epsilon^{\!2}}\varphi+\dots\,.
\end{align}
Here's the argument: in expanding  Eq.~\eqref{ep3} according to
formula \eqref{mu1} we can omit to explicitly writing the cosines,
since we know that each $e,E$ multiplies $\cos
k_{\epsilon^{\!2}}\varphi$, so as each $B$ and $C$ multiplies $\cos
2k_{\epsilon^{\!2}}\varphi$ and $\cos 3k_{\epsilon^{\!2}}\varphi$
respectively. This way we can proceed rapidly by inspecting more
concise expressions. Let us filter now Eq.~\eqref{ep3} through the
sieve represented by the constraints we have imposed to isolate the
resonant term $\sim\epsilon^3\cos k_{\epsilon^{\!2}}\varphi$.

We observe first that the squared terms in Eq.~\eqref{mu1} can be
deleted because of Eq.~\eqref{mu2} for $n=1,2,3$. Next, we drop all
double products of Eq.~\eqref{mu1} in which the sum of the subscript
indices is different from 2 and so, after multiplication by
$\epsilon$, will survive only the terms of order $\epsilon^3$. Last,
when we meet products of two cosines, we consider only those in
which the arguments differ by one, and apply to them
Eq.~\eqref{mu3}. At the end we obtain the following sum of
coefficients of $\cos k_{\epsilon^{\!2}}\varphi$
\begin{align}
2\epsilon eA_{2}+\epsilon eB_{2}&+2\epsilon A_{1}E_{1}+\epsilon
E_{1}B_{1},
\end{align}
but the last two terms are erased by the $\epsilon^3$-coefficients
arising from the rightmost expression of Eq.~\eqref{did}, which are
the constants $-2\epsilon A_{1}$ and $-\epsilon B_{1}$ times the
term $E_{1}\cos k_{\epsilon^{\!2}}\varphi$ of $w_{2}$, and thus we
finally get
\begin{equation}
H_{3}=\epsilon e(2A_{2}+B_{2}).
\end{equation}


\begin{thebibliography}{}


\bibitem[Bertotti(2003)]{ber03}
Bertotti B., Farinella P., Vokrouhlick´y D., 2003. \emph{Physics of
the Solar System}. (Dordrecht: Kluwer). p. 313.

\bibitem[Burgay(2003)]{bur03}
Burgay, M.; D'Amico, N.; Possenti, A.; Manchester, R. N.; Lyne, A.
G.; Joshi, B. C.; McLaughlin, M. A.; Kramer, M.; Sarkissian, J. M.;
Camilo, F.; Kalogera, V.; Kim, C.; Lorimer, D. R. \emph{An increased
estimate of the merger rate of double neutron stars from
observations of a highly relativistic system}.Nature, Volume 426,
Issue 6966, pp. 531-533 (2003).

\bibitem[Calura(1997)]{cal97}Calura, M.; Fortini, P.; Montanari, E.
\emph{Post-Newtonian Lagrangian planetary equations}, Physical
Review D (Particles, Fields, Gravitation, and Cosmology), Volume 56,
Issue 8, 15 October 1997, pp.4782-4788

\bibitem[d'Alembert(1754)]{dal54} d'Alembert, A. \emph{Recherches sur Differents
Points du Systeme du Monde}. (1754-56) Livre 1 Chap. VI, n.27.

\bibitem[Damour(1988)]{dam88} Damour, T.,  Schaefer, G.  \emph{Higher-order relativistic periastron
advances and binary pulsars},  Nuovo Cimento B, Volume 101, No. 2.

\bibitem[D'Eliseo(2007)]{del07} D'Eliseo, M.M. 2007  \emph{The first-order orbital equation},
Am. J. Phys. \textbf{75}, 352

\bibitem[d'Inverno(1992)]{din92} d'Inverno, A. \emph{Introducing Einstein's Relativity}, Oxford Univerity Press,
Oxford, 1992.

\bibitem[Do-Nhat(1998)]{don98} Do-Nhat T. \emph{Full asymptotic expansion of the relativistic orbit of a test
particle under the exact Schwarzschild metric}. Physics Letters A
238 ( 1998) 328-336.

\bibitem[Iorio(2006)]{ior06}
Iorio, L. \emph{Are we far from testing general relativity with the
transiting extrasolar planet HD 209458b 'Osiris'?} New Astronomy,
Volume 11, Issue 7, p. 490-494, 2006.

\bibitem[Iorio(2007)]{ior07} Iorio, L. \emph{The post-Newtonian mean anomaly advance as further
post-Keplerian parameter in pulsar binary systems}, Astrophys. Space
Sci., Volume 312, Numbers 3-4. December, 2007, pp. 331-335.

\bibitem[Lin-Sen Li(2010)]{lin10} Lin-Sen Li, \emph{Post-Newtonian effect on the variation of time of
periastron passage of binary stars in three gravitational theories},
Astrophys. Space Sci., doi:10.1007/s10509-010-0267-4

\bibitem[Kramer(2006)]{kra06}
Kramer, M.; Stairs, I. H.; Manchester, R. N.; McLaughlin, M. A.;
Lyne, A. G.; Ferdman, R. D.; Burgay, M.; Lorimer, D. R.; Possenti,
A.; D'Amico, N.; Sarkissian, J. M.; Hobbs, G. B.; Reynolds, J. E.;
Freire, P. C. C.; Camilo, F. \emph{Tests of General Relativity from
Timing the Double Pulsar}, Science, Volume 314, Issue 5796, pp.
97-102 (2006).


\bibitem[Landau(1966)]{lan66} Landau, L. - Lifchitz, E. \emph{Th\'{e}orie du Champ},  Editions MIR
Moscou, 1966, p.415.

\bibitem[Lindstedt(1883)]{lin83} Lindstedt, A. \emph{Astr.Nachr. Nr 2482}, 1883.

\bibitem[Mason(1984)]{mas84} Mason, D.P. - Wright, C.J., \emph{An Improved Singular
Perturbation Solution for Bound Geodesic Orbits in the Schwarzschild
Metric}. General Relativity and Gravitation, 17ol. 16, No. 2, 1984.

\bibitem[Nayfeh(1973)]{nay73} Nayfeh, A.H. \emph{Perturbation Methods}, Interscience,
New York, 1973.

\bibitem[pal(2008)]{pal08}
P\`{a}l, A.; Kocsis, B., \emph{Periastron precession measurements in
transiting extrasolar planetary systems at the level of general
relativity}, Monthly Notices of the Royal Astronomical Society,
Volume 389, Issue 1, pp. 191-198. 2008.


\bibitem[Schafer(1993)]{sch93}
Schafer, G., Wex, N. Phys. Lett. A 174 (1993) 196; 177 (1993) 461
(E).

\end{thebibliography}
\end{document}